\RequirePackage{fix-cm}
\documentclass[smallextended]{svjour3}       % onecolumn (second format)
\smartqed  % flush right qed marks, e.g. at end of proof
\usepackage{graphicx}
\usepackage{hyperref}
\usepackage{siunitx}
\usepackage{xcolor}
\usepackage{listings}

\lstdefinestyle{mystyle}{
    basicstyle=\ttfamily\footnotesize,
    breakatwhitespace=false,         
    breaklines=true,                 
    captionpos=b,                    
    keepspaces=true,                 
    numbers=left,                    
    numbersep=5pt,                  
    showspaces=false,                
    showstringspaces=false,
    showtabs=false,                  
    tabsize=2,
    frame=single
}

\newcommand\YAMLcolonstyle{\color{red}\mdseries}
\newcommand\YAMLkeystyle{\color{black}\bfseries\footnotesize}
\newcommand\YAMLvaluestyle{\color{blue}\mdseries}

\makeatletter

\newcommand\language@yaml{yaml}

\expandafter\expandafter\expandafter\lstdefinelanguage
\expandafter{\language@yaml}
{
  keywords={true,false,null,y,n},
  keywordstyle=\color{darkgray}\bfseries,
  basicstyle=\YAMLkeystyle,                                 % assuming a key comes first
  sensitive=false,
  comment=[l]{\#},
  morecomment=[s]{/*}{*/},
  commentstyle=\color{purple}\ttfamily,
  stringstyle=\YAMLvaluestyle\ttfamily,
  moredelim=[l][\color{orange}]{\&},
  moredelim=[l][\color{magenta}]{*},
  moredelim=**[il][\YAMLcolonstyle{:}\YAMLvaluestyle]{:},   % switch to value style at :
  morestring=[b]',
  morestring=[b]",
  literate =    {---}{{\ProcessThreeDashes}}3
                {>}{{\textcolor{red}\textgreater}}1     
                {|}{{\textcolor{red}\textbar}}1 
                {\ -\ }{{\mdseries\ -\ }}3,
}

\lst@AddToHook{EveryLine}{\ifx\lst@language\language@yaml\YAMLkeystyle\fi}
\makeatother

\newcommand\ProcessThreeDashes{\llap{\color{cyan}\mdseries-{-}-}}
\newcommand{\SD}{{\em SecDocker }}

\begin{document}

% -------------------------------------------------------------------

\title{%
     SecDocker: Hardening the Continuous Integration Workflow
    %Proposal1: Cybersecurity Challenges in Continuous Integration Process
    
}
\subtitle{Wrapping the container layer}

\author{%
    David Fernández González \and
    Francisco Javier Rodríguez Lera \and 
    Gonzalo Esteban \and
    Camino Fernández Llamas
}


\institute{%
    David Fernández González \at
    Campus de Vegazana, Universidad de León \\
    %   Tel.: +123-45-678910\\
    %   Fax: +123-45-678910\\
    \email{dferng@unileon.es}           %  \\
        %   \and
        %   S. Author \at
        %       second address
}

\date{Received: date / Accepted: date}

% -------------------------------------------------------------------

\maketitle

% -------------------------------------------------------------------

\begin{abstract}

Current Continuous Integration processes face significant intrinsic cybersecurity challenges. The idea is not only to solve and test formal or regulatory security requirements of source code but also to adhere to the same principles to the CI pipeline itself. This paper presents an overview of current security issues in CI workflow. It designs, develops, and deploys a new tool for the secure deployment of a container-based CI pipeline flow without slowing down release cycles. The tool, called \SD for its Docker-based approach, is publicly available in GitHub. It implements a transparent application firewall based on a configuration mechanism avoiding issues in the CI workflow associated with intended or unintended container configurations. Integrated with other DevOps Engineers tools, it provides feedback from only those scenarios that match specific patterns, addressing future container security issues.


\end{abstract}

% -------------------------------------------------------------------
\section{Introduction}
\label{sec:introduction}

%   Justification I
There are plenty of tools to analyze and secure the creation of
container images.
Besides that,
several organizations have developed guidelines to assist developers
in the creation of such images with a certain degree of security.
For instance,
focusing on Docker~\cite{Martin:2018},
it is possible to find the \emph{Docker Benchmark tool} released
by the Center for Internet Security
(CIS)~\cite{Goyal:2017} or the \emph{Ultimate Benchmark for
Container Image Scanning}
(UBCIS)~\cite{Berkovich:2020};
both containing guides that analyze every single dangerous step
involved during the image-building process.

%   Justification II
However,
there are different local exploitation issues associated to the
Continuous Integration
(CI)
workflow that needs to be secured.
The problem here is that some containerization solutions---like
Docker---has different exploits that allow an attacker to override
some of the image specifications;
which is done by providing new ones at the very moment the
the container is created.
Furthermore,
opening ports is always a security risk if it is controlled by a low
level user in the system
(like a developer in a DevOps server).

%   Motivation
In this way,
this study focuses on developing a tool called \SD for
enhancing the cybersecurity pipeline when integrating containerization
in the CI workflow~\cite{Vase:2016}.
\SD is a wrapper,
specifically an application firewall for Docker,
that allows the sysadmins to block the capabilities offered by Docker
in the \texttt{run} command.
By doing so,
even if Docker allows the user to perform dangerous activities the
action is blocked before it gets executed.

Below,
the research question and main contribution are presented.
The remainder of the paper presents the elements for answering the
research question.
Section~\ref{sec:background} overviews the state of the art in
containerization and Continuous Integration workflow. Section~\ref{sec:containers_ci} presents the developer's and attacker's
scheme to the containerized CI layer.
Section~\ref{sec:solution} presents \SD tool:
its design,
architecture and usage.
Section~\ref{sec:validation} validates with the results of \SD from
two different experiments carried out in this research measuring performance and the operational flow.
Section~\ref{sec:discussion} discusses enumerating pro's and cons of
\SD and finally Section~\ref{sec:conclusions} provides conclusions
about the processes and solutions presented in this paper. 

\subsection{Research Question and Contribution}%
\label{subsec:introduction:rq}

%   TODO: review this section
Continuous Integration is a cornerstone methodology that addresses
automatically several processes previously faced by software developers.
However, 
the CI workflow also needs to meet the security mechanisms that will
guarantee flexibility,
productivity,
and efficiency during the software development life cycle.
Thus,
this paper aims a set of elements that are framing within the next
Research Question
(RQ):

\begin{description}
    \item[RQ:] \emph{Which are the mechanisms for avoiding and minimizing
    cybersecurity and misconfiguration issues in a CI container-based
    deployment system?}
\end{description}

This RQ scales to a new level when the containerization tool is Docker. Most parts of current automation servers and processes are supported on Docker containers. However, its engine shows critical points that favor malicious users or an unaware DevOps engineer to crash the CI pipeline. 
Working with an erroneous configuration would promote a bad system behavior, with the manpower and economical costs associated. There are three main phases when using Docker containers in the CI workflow: 1) issues associated with image retriever; 2) issues associated with image builder, and 3) issues generated when the image is deployed. 

This study presents the overview of the latest step, as well as the design and development of a tool called \SD for minimizing issues associated with container deployment. Besides, the tool introduces expansion capabilities for solving the first and second steps using plugins.

% What is the Role of our tool SecDocker in the CI?

% -------------------------------------------------------------------

\section{Background}%
\label{sec:background}

%   What is CI, CDE, and CD
CI is one of many software development practices aimed at helping organizations
to accelerate their development and delivery of software features without
compromising quality~\cite{Humble:2010}.
According to Fitzgerald and Stol~\cite{Fitzgerald:2017},
it can be defined as ``a process which is typically automatically triggered
and comprises inter-connected steps such as compiling code,
running unit and acceptance tests,
validating code coverage,
checking compliance with coding standards and building deployment
packages''.
For Shahin, Babar, and Zhu~\cite{Shahin:2017},
alongside Continuous Delivery or CDE
(ensure the package is always at a production-ready state after tests)
and Continuous Deployment or CD,
(deploy the package to production or customer environments),
CI is considered part of the continuous software engineering paradigm,
which includes the popular term ``DevOps''~\cite{Fitzgerald:2017}.

%   What is DevOps and its tools
DevOps is a mix of the words \emph{Development} and \emph{Operations} and,
although there is no common definition for it,
some literature reviews exist to date that addresses this
point~\cite{Smeds:2015,Jabbari:2016,Ghantous:2017}.
For instance,
Jabbari et al.~\cite{Jabbari:2016} define it as
``a development methodology aimed at bridging the gap between Development
(Dev)
and Operations
(Ops),
emphasizing communication and collaboration,
continuous integration,
quality assurance and delivery with automated deployment utilizing a set of
development practices''.
To enable such concepts or practices,
and thus aid developers in materializing them,
DevOps relies on using a range of tools~\cite{Ghantous:2017,Leite:2019};
from source code management to monitoring and logging,
as well as configuration management.
Together,
these tools allow the creation of a pipeline that automate the processes
of compiling,
building and deploying the source code into a production
platform~\cite{Humble:2010}.

%   DevOps issues related to the pipelines and approaches
But as a relative young methodology,
integrating and maintaining these tools or managing the infrastructure
in which they run automatically may pose a challenge~\cite{Shahin:2017};
especially for CI and CD.
As Leite et al. discuss in their literature review~\cite{Leite:2019},
concepts like ``infrastructure as code'',
virtualization,
containerization or cloud services are solutions currently known to
be used for these types of issues.
Among all of them,
containerization is perhaps the most popular solution in DevOps
environments at the moment.
With a 
platform as a service focus,
it is used for delivering software in a portable and streamlining
the way by providing a platform that allows  developing,
running and managing applications without worrying about the infrastructure
needed~\cite{Pahl:2015}.
% https://www.ibm.com/downloads/cas/BBKLLK1L | IBM report on cloud computing (containers usage)

%   Containers: definition and features
Technically speaking,
containerization is a type of lightweight OS-level virtualization
technology that allows running multiple isolated systems
(in terms of processes,
resources,
network,
etc)
while sharing the same host OS.
Such systems or \emph{containers},
hold packaged,
self-contained applications and,
if necessary,
binaries and libraries required to run them~\cite{Bernstein:2014}.
Moreover,
they have been around for some time in various forms:
from chroot,
FreeBSD jails or Solaris zones to Linux-based solutions relying on
kernel support like LXC or
OpenVZ~\cite{Bernstein:2014,Turnbull:2014,Pahl:2015,Combe:2016}.
But over time,
containerization has become a major trend thanks to tools like
Docker~\cite{Merkel:2014,Leite:2019}.

%   Docker briefly overview
Docker is an open-source platform that facilitates the management
of containers by using a client-server architecture through a
CLI tool,
a daemon and a REST API~\cite{Merkel:2014,Turnbull:2014}.
It relies on the concept of \emph{images} to build containers,
that is,
a specification of the collection of layered file systems,
their corresponding execution environment and some metadata;
making them portable,
shareable and also updatable~\cite{Pahl:2015}.
Regarding their usage,
Docker containers can be used either as a microservice
(to host a single service),
as a way of shipping complete virtual environments
(to reproduce and automate the deployment of applications)
or even as a platform as a service
(to cope with security and infrastructure integration
issues)~\cite{Combe:2016,Martin:2018}.

%   Security issues on Docker and how to address them
From a security perspective,
Docker provides different levels of isolation,
host hardening capabilities and some countermeasures related
to network operations~\cite{Chelladhurai:2016,Combe:2016,Martin:2018}.
Nevertheless,
it is not exempt from security threats nor vulnerabilities
such as ARP spoofing,
DoS attacks,
privilege escalation,
etc.
This is due to the nature of containerization itself because an
attack on the host OS may expose all containers and their
network traffic.
To address these cybersecurity risks it is necessary to take
similar actions to DevOps;
especially where pipeline automation is a requirement
(as in CI or CD).
Such actions can be understood as best practices or recommendations
that aim to establish a Secure Software Development Life Cycle.
Examples of this can be found in reports like \emph{DevSecOps:
How to Seamlessly Integrate Security Into DevOps}~\cite{MacDonald:2016}
or \emph{DoD Enterprise DevSecOps Reference Design}~\cite{CIO:2019},
where container hardening is contemplated.

% -------------------------------------------------------------------

\section{Container Layer in CI}%
\label{sec:containers_ci}

Containers are used in CI processes to isolate and automate the creation
of an application into one single self-contained virtual environment.
This solution simplifies DevOps manpower,
as it allows to split a large application development project into several
smaller work units.
Having said that,
this section describes the role of CI from the point of view of two actors
(DevOps engineers and attackers)
and also presents the scenarios likely to be vulnerable.

\subsection{DevOps Engineers Scheme}
\label{subsec:containers_ci:devops_eng}

%   Role of containers in CI
From a developers perspective,
CI is used to guarantee the quality,
consistency and viability across different environments~\cite{Hilton:2017}.
But as CI systems are vulnerable to security attacks and
misconfigurations~\cite{Shahin:2017},
DevOps engineers frequently rely on containers to create such
environments as they provide isolation without much effort to
them.
Generally,
this has been achieved by technologies like Docker,
which allow them to treat infrastructures as
code~\cite{Kang:2016}.

%   Role of Docker in CI
Regarding CI,
Docker has ease DevOps engineers in the replication of
environments for building automation pipelines.
Particularly,
as Boettinger et al. point in their work~\cite{Boettinger:2015},
it has solved common issues encountered by end-users like managing
dependencies
(through images),
imprecise documentation
(through scripts to build up such images)
or code-rot
(with image versioning);
along with the adoption and re-use of existing workflows
(thanks to features like portability,
easy integration into local environments or public repositories
for sharing and reusing those images).

%   Role of DevOps engineers in CI
But despite the benefits that Docker or other containerization
technologies may offer to DevOps engineers in CI environments,
the latter still face challenges related to its adoption;
particularly associated with introducing any new technologies or phenomena in a given organization~\cite{Hilton:2017,Shahin:2017}.
According to Shahin et al.~\cite{Shahin:2017},
literature shows that,
among the common practices for implementing CI workflows,
DevOps engineers need to decompose development into smaller
units and also plan and document the activities that comprise
the automation pipeline.
Having said that,
it must be noted that there are many ways of approaching the
design of such pipelines.
But taking into account the use of containers and based on
Bass et al. approach~\cite{Bass:2015},
any CI workflow must include the following 6 components in
such design plan:

\begin{enumerate}

    \item Automation server.
    Implements the CI/CD pipeline and creates a local workspace
    in which its steps take place.
 
    \item Orchestrator.
    Sequentially triggers each step of the pipeline by communicating
    with the remaining components.
    It should be noted that,
    when using containers,
    steps may require images to perform their actions.
    Thus,
    the same image can be used through the whole pipeline or in
    specific steps.

    \item Code retriever.
    Pulls source code from repository to local workspace.

    \item Unit tester.
    Runs automated unit tests on source code.

    \item Artifact builder.
    Builds deployable artifacts from source code.

    \item Image generator.
    Builds,
    verifies,
    stores,
    and deploys an image to be used within the pipeline.
    %testing/production environments.

\end{enumerate}

%Making a standardized process to establish and defined this infrastructure will lower security and functional risk. Different automated continuous tests could are applied to the deployment process, however particularly for item 6, some of the tools go toward a specified commercial solution. Thus, it is justified the development of \SD.
With this in mind and despite using containers,
any standard CI workflow that establishes and defines this
components will lower security and increase its functionality risks.
To avoid this,
different automated continuous tests could be applied to the whole
process.
However,
and particularly for item 6,
some of the tools go toward a specific commercial solution.
As a result,
there is a need to develop a tool like \SD.

\subsection{Attacker Scheme}
\label{subsec:containers_ci:attackers}

%   Introduction
As mentioned in Section~\ref{sec:background},
containers are the target of different security threats
or vulnerabilities.
Therefore,
a containerized environment---like those created with
Docker---may have different potential attack
vectors~\cite{Martin:2018}:
host OS,
network or physical systems,
source code repositories,
image repositories or the very own containers.
Securing these vectors is not a trivial task,
but the contributions presented in this paper are framed
towards the integrity of container images used by CI
(or CD)
pipelines.

\begin{figure}[ht]
     \centering
     \includegraphics[width=\linewidth]{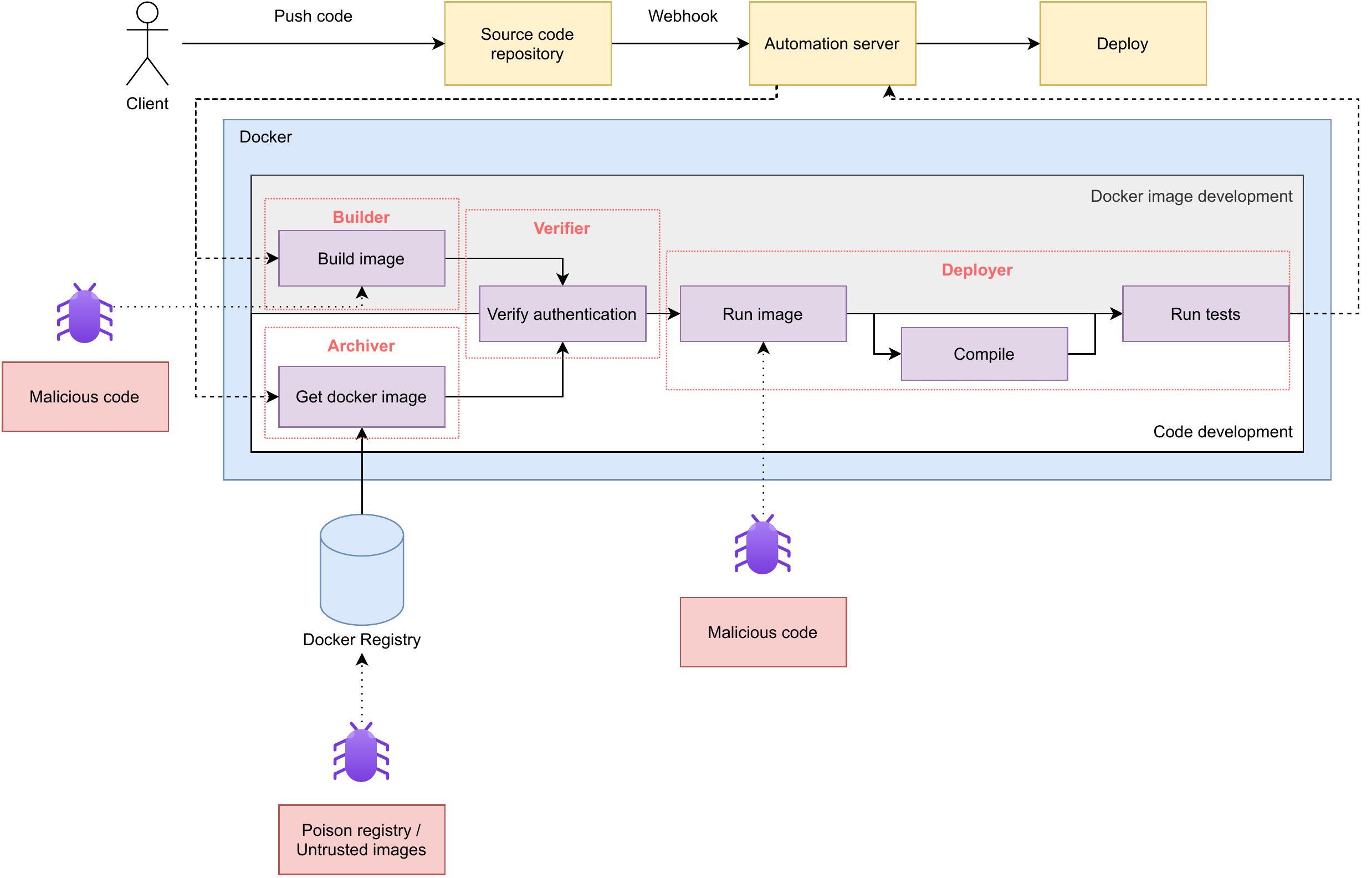}
     \caption{Attacker scheme: Vulnerability points during container deployment.}
     \label{fig:Containerissues}
\end{figure}

%   Image hardening
In such cases,
images are frequently used to ship a complete virtual environment where concrete actions from the CI workflow
take place
(e.g. build,
test,
run or deploy an application).
Such workflow is scripted and usually automated by triggering
a webhook from some version control system.
But this approach makes pipelines unreliable so,
to contribute to its hardening,
the image generator component from the CI process
(see the previous subsection)
is an element that needs to be hardened somehow.
Regarding this process and based on Bass et al.
approach~\cite{Bass:2015},
it is possible to distinguish four components involved in it (Figure~\ref{fig:Containerissues}):

\begin{enumerate}

    \item Builder.
    Builds a container image according to some specifications.
    This image comprises the virtual environment or workspace 
    where some or all workflow actions will take place.

    \item Verifier.
    Computes a checksum in order to verify the authentication of
    the image was just built.

    \item Archiver.
    Stores the image in a registry or repository so it can be
    retrieved later.

    \item Deployer.
    Deploys the image into a testing or production environment in
    order to execute the CI workflow or some of its scripted actions.

\end{enumerate}

This study considers the last component to be one of the most important.
The reason is that a correct configuration will minimize the impact of
an issue in the previous three components.
A container with no root or bounded CPU will guarantee minimal resource
exploitation to the host machine.
Thus,
a runtime check for the detection of common security and configuration
weaknesses against a compliance configuration pattern defined by DevOps
engineers seems to meet the requirements for production environments.

% -------------------------------------------------------------------

\section{Proposed Solution}
\label{sec:solution}

%   SecDocker definition
\SD is an application firewall for Docker and thus shares the
same purpose as any web application firewall:
prevent users from performing dangerous or unexpected actions on
systems.

\begin{figure}[ht]
    \centering
    \includegraphics[width=\linewidth]{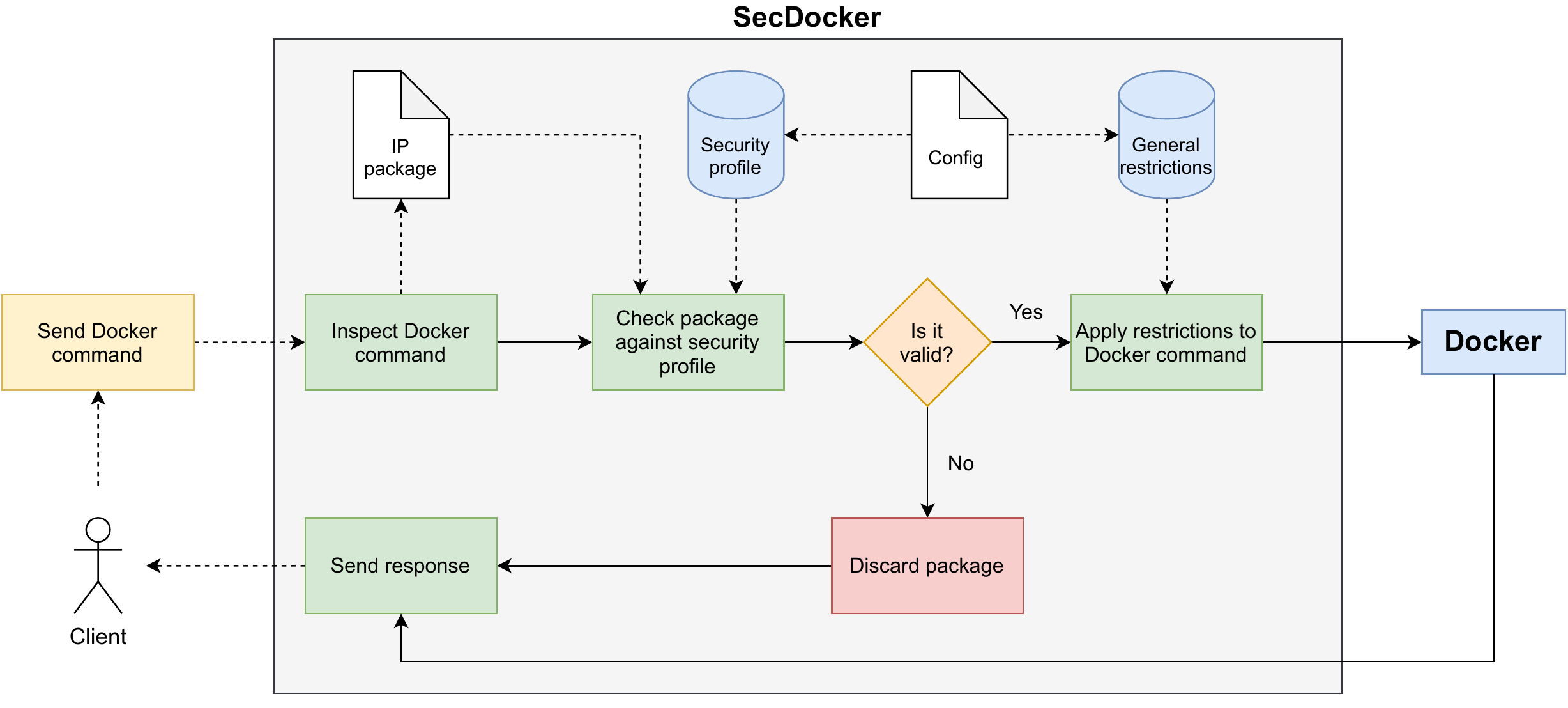}
    \caption{\SD workflow}
    \label{fig:solution:workflow}
\end{figure}

%   Workflow description
Broadly speaking, 
the solution filters TCP traffic and works by monitoring the Docker
run command. Its main goal is to evaluate all the requests meant for the Docker
daemon by standing between it and the user
(see Figure~\ref{fig:solution:workflow}).

Whenever a new HTTP request aiming for the run API endpoint reaches
the firewall,
the IP package is opened and the request parameters are loaded.
After that,
these variables are checked against a security profile
(previously configured by the user)
to prevent unauthorized actions.
On the one hand,
when a match is found that package is dropped and a new one is created
and sent to the end-user.
It should be noted that this ``new'' package contains the original data
plus the information that one or more requested options were not allowed.
On the other hand,---when no matches are found---,
\SD appends or modifies the requested parameters to suit some
running restrictions
(previously specified by the user).
Then,
the package is recreated and sent to the Docker daemon to finally
perform the requested action.

\subsection{Software architecture}
\label{subsec:solution:architecture}

%   Core modules
\SD is written in Go and is publicly available
in GitHub\footnote{https://github.com/uleroboticsgroup/Secdocker}.
It features a modular and extensible design comprised of 5
components at its core:

\begin{enumerate}

    \item \emph{Security}.
    Performs validation against the user-supplied options.
    
    \item \emph{Config}.
    Loads user's information into the firewall in real-time.
    
    \item \emph{Docker}.
    Performs tasks related to how Docker processes information.
    %   TODO: insert some examples?
    
    \item \emph{HTTPServer}.
    Manages and performs actions against HTTP data
    (e.g. loading the body of the requests,
    crafting new requests/responses,
    etc).
    
    \item \emph{TCPIntercept}.
    Handles packages at the TCP level,
    so the communication looks transparent for the end-user.
    It also maintains the communications and gathers data for the
    \texttt{HTTPServer} module.
    Additionally,
    it must be noted that this module is based on
    Trudy\footnote{\url{https://github.com/praetorian-inc/trudy}},
    %~\cite{Trudy:2021},
    a transparent proxy that can modify and drop TCP traffic.

\end{enumerate}

%   Plugins module
Its functionality can also be expanded by third-party applications
thanks to a dedicated component named \texttt{Plugins}.
For its basic workflow,
\SD delegates some extra functionality to two plugins:

\begin{itemize}

    %~\cite{Anchore:2021}  %https://github.com/anchore/anchore-engine
    \item Anchore\footnote{\url{https://anchore.com/}}.
    Inspects,
    analyzes and applies user-defined acceptance policies.
    
    %~\cite{Notary:2021}. %https://github.com/theupdateframework/notary
    \item Notary\footnote{\url{https://docs.docker.com/notary/getting_started/}}       Ensures the integrity of a trusted collection of Docker images.
    
\end{itemize}

Likewise,
an accountability component based on logs is also included with
\SD.
This logging component relies on
Logrus\footnote{\url{https://github.com/sirupsen/logrus/}},
an external logger package for Go that provides structured logs.

\subsection{Usage}
\label{subsec:solution:applications}

\begin{figure}[ht]
    \centering
    \includegraphics[width=\linewidth]{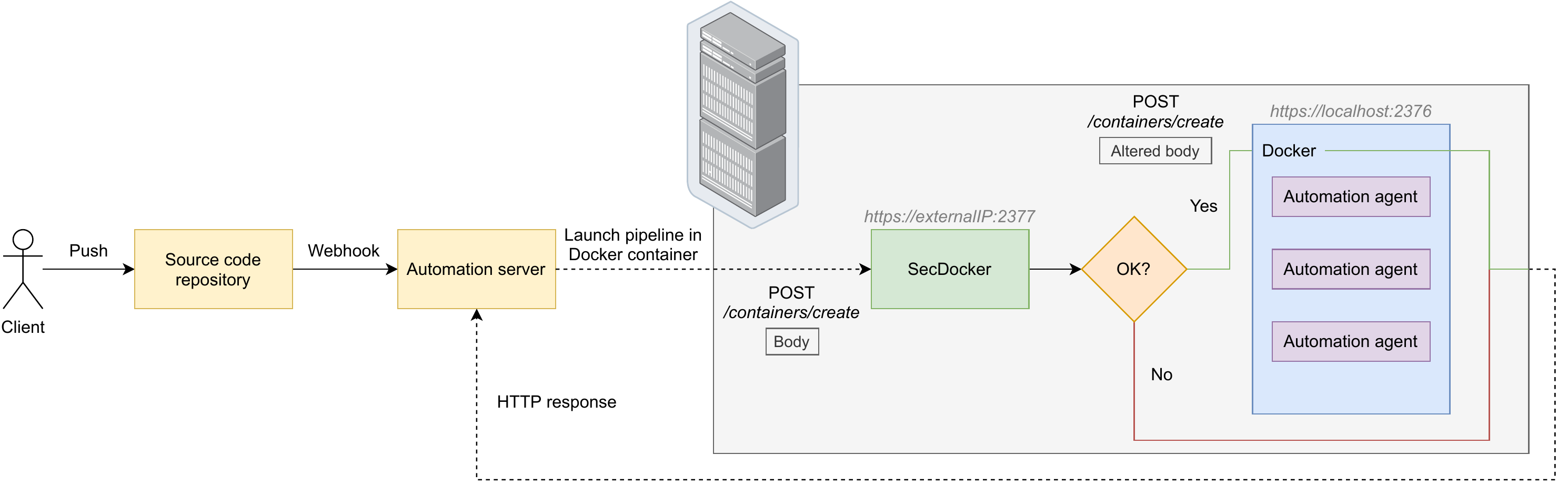}
    \caption{\SD general overview}
    \label{fig:solution:overview}
\end{figure}

%   Basic usage
As mentioned at the beginning of this section,
\SD workflow involves routing TCP packages in a similar way
to a firewall.
Consequently,
it should be placed in front of the server responsible for handling
requests to Docker.
This is done either to maintain the original destination port of
the Docker daemon or to perform some alteration to redirect
the traffic to the right port.
Figure~\ref{fig:solution:overview} illustrates the process and places
\SD in the CI workflow.

Furthermore,
to filter all this traffic a YAML file is used
(see Listing~\ref{lst:configuration} for an example).
This file contains a set of configurable features that allow end
users to set up some security features related to the Docker image
and its execution
(see Table~\ref{tab:solution:features}).

\begin{lstlisting}[%
    language=yaml,
    float,
    caption={Example of a \SD configuration file},
    label={lst:configuration}%
]
plugins:

dockerapi: "/var/run/docker.sock"

restrictions:
  ports:
    - 22
    - 25
  mounts:
    - /root
    - /
  users:
    - root
  environment:
    - USER=0
  images:
    - ubuntu:16.04
  privileged: true

general:
  memory: "1g"
  cpu: 0.25
  user: "1000"
  environment:
    - "MY_ENV=true"
\end{lstlisting}

\begin{table}[ht]
    \centering
    \caption{Configurable features supported by \SD}
    \label{tab:solution:features}
    %%%%%%%%%%%%%%%%%%%%%%%%%%%%%%%%%%%%%%%%%%%%%%%%%%%%%%%%%%%%%%%%%%%%%
%
%   tables/solution-secdocker.tex
%
%%%%%%%%%%%%%%%%%%%%%%%%%%%%%%%%%%%%%%%%%%%%%%%%%%%%%%%%%%%%%%%%%%%%%

\begin{tabular}{p{.22\linewidth} l p{.49\linewidth}}
\hline\noalign{\smallskip}
% -------------------------------------------------------------------
Feature type
    & Setting
        & Description\\
% -------------------------------------------------------------------
\noalign{\smallskip}\hline\noalign{\smallskip}
Docker options
    & Ports
        & Port number(s) used by the container\\
    & Users
        & Username(s) or uid(s) running the container\\
    & Mounts
        & Mounted volume(s)
        (file or directory)
        from the host filesystem\\
    & Environment
        & List of environment variables
        (KEY=VAL)
        used within the container\\
    & Security Policies
        & Adds or drops Linux capabilities\\
    & Images
        & Name(s) of container image(s) to monitor\\
    & Privileged
        %& Grants a container the capabilities of its host machine\\
        & Whether the containers has granted capabilities in its host machine\\
% -------------------------------------------------------------------
General restrictions
    & Memory
        & Maximum amount of memory usage\\
    & CPU
        & Number of CPU resources\\
    & User
        & Username or uid
        (in host)
        allowed to run the container\\
    & Environment
        & Environment variables of the container\\
% -------------------------------------------------------------------
\noalign{\smallskip}\hline
\end{tabular}
\end{table}

%   Output
Regarding its output,
\SD starts to listen on port 8999 and logs all packages and
their related data to the standard output by default.
A separated log file is also created containing all the requests;
whether they were allowed or not and why.
Furthermore,
any external plugins can have their logs to output their own
results.

% -------------------------------------------------------------------

\section{Validation}%
\label{sec:validation}

%   Setup description
Two experiments were carried out to both measure \SD's
performance and check its functionality.
The experiments were conducted on two PCs connected to the same LAN.
Both systems were configured using Elementary OS 5.1.7 and had
different specifications:
one with an Intel(R) i5-3570 CPU @ 3.40 GHz and 16.0 GB memory,
and the other with an AMD Ryzen 5 3500U CPU @ 3.60 GHz and 8.0 GB
memory.
The first PC was used as a server for running Docker and \SD
and the second as a client to connect to the latter and execute
different Docker commands.

\subsection{Performance Testing}%
\label{subsec:experiment:performance}

%   Experiment goal
The first experiment was carried out to evaluate \SD's performance
as timing behavior,
that is,
to run transparently from a running Docker server.
The test consisted of running 100 times the following command
from the client's PC:
\begin{equation}
    \texttt{\# docker run alpine:latest}
\end{equation}
No extra arguments were provided to ensure that no external factors
(like an increase of CPU usage or network speed)
affects the measurements.
Besides,
the standard Unix \texttt{time} tool was used to measure the performance
of each command.

%   Experiment results
Table~\ref{tab:solution:performance} summarizes the experiment
results after its execution with and without \SD.
The mean time for running commands when \SD was enabled
in the server PC was $0.509\pm{}0.085$ seconds,
with an interquartile range of $0.06$ seconds.
Meanwhile,
the meantime for commands when \SD was disabled in the
the same server was $0.519\pm{}0.041$ seconds,
with an interquartile range of $0.05$ seconds.
Since the differences are only $0.01$ seconds between the mean
times and also between the interquartile ranges,
it can be considered valid to assert that,
apparently,
\SD runs transparently from Docker.

\begin{table}[ht]
    \centering
    \caption{Statistics related to time taken to execute 100 Docker
    commands when \SD is both enabled and disabled}
    \label{tab:solution:performance}
    %%%%%%%%%%%%%%%%%%%%%%%%%%%%%%%%%%%%%%%%%%%%%%%%%%%%%%%%%%%%%%%%%%%%%
%
%   tables/experiment-performance.tex
%
%%%%%%%%%%%%%%%%%%%%%%%%%%%%%%%%%%%%%%%%%%%%%%%%%%%%%%%%%%%%%%%%%%%%%

\begin{tabular}{r c c}
\hline\noalign{\smallskip}
% -------------------------------------------------------------------
Statistics
    & Without SecDocker
        & With SecDocker\\
    & (secs)
        & (secs)\\
% -------------------------------------------------------------------
\noalign{\smallskip}\hline\noalign{\smallskip}
Mean
    & \num[round-mode=places,round-precision=3]{0.518585858585859}
        & \num[round-mode=places,round-precision=3]{0.509393939393939}\\
Standard Error
    & \num[round-mode=places,round-precision=3]{0.004116450901859}
        & \num[round-mode=places,round-precision=3]{0.008538679841084}\\
Mode
    & 0.540
        & 0.540\\
Median
    & 0.520
        & 0.510\\
First Quartile
    & 0.490
        & 0.480\\
Third Quartile
    & 0.540
        & 0.540\\
Variance
    & \num[round-mode=places,round-precision=3]{0.001677571634715}
        & \num[round-mode=places,round-precision=3]{0.007217996289425}\\
Standard Deviation
    & \num[round-mode=places,round-precision=3]{0.040958169328163}
        & \num[round-mode=places,round-precision=3]{0.084958791713541}\\
% -------------------------------------------------------------------
\noalign{\smallskip}\hline
\end{tabular}
\end{table}

\subsection{Functional Testing}
\label{subsec:experiment:functional}

%   Experiment goal
The second experiment was carried out to test \SD functionality.
This time,
the goal was to send the following command from the client's PC
to perform a hypothetical privilege escalation attack:
\begin{equation}
    \texttt{\# docker run --privileged ubuntu:18.04}
\end{equation}
To prevent this potential threat,
the server PC used the same configuration file shown in
Listing~\ref{lst:configuration};
which includes the \texttt{privileged} option set to \emph{true} in
order to drop commands like the one previously mentioned.

\begin{figure}[ht]
    \centering
    \includegraphics[width=\linewidth]{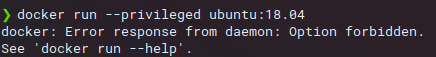}
    \caption{\SD response when running the proposed Docker command}
    \label{fig:solution:forbidden}
\end{figure}

\begin{figure}[ht]
    \centering
    \includegraphics[width=\linewidth]{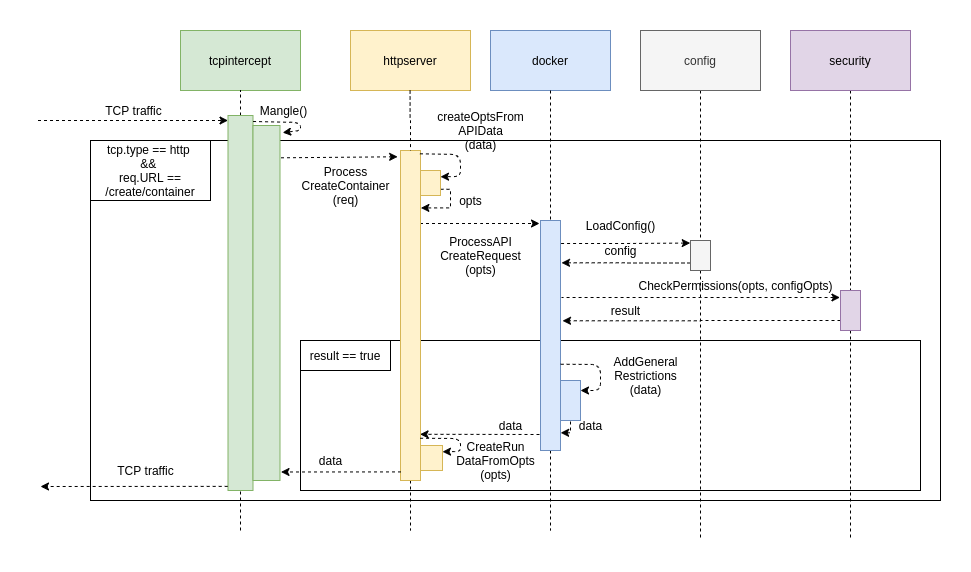}
    \caption{Sequence diagram describing how \SD blocks a
    \texttt{docker run} command}
    \label{fig:solution:seqdiagram}
\end{figure}

%   Experiment results
Figure~\ref{fig:solution:forbidden} shows that running the proposed
command for this test fails as expected.
From \SD's point of view,
the command is processed as represented in the sequence diagram shown
in Figure~\ref{fig:solution:seqdiagram}.
When the HTTP request derived from the command arrives at \SD,
it extracts all parameters and checks them against the security
configuration loaded.
In the test environment, the privileged option is met,
so a response is sent to the user stating that it has a forbidden option.

% -------------------------------------------------------------------

\section{Discussion}
\label{sec:discussion}

\SD is a tool meant to be used for a wide variety of members from
CI and DevOps communities.

\subsection{Research Perspective}
\label{subsec:discussion:research_perspective}

This work makes the assumption that the RQ proposed in 
Section~\ref{subsec:introduction:rq}
(see below)
is appropriate,
meaningful,
and purposeful when facing cybersecurity issues during the CI workflow.

\begin{description}
    \item[RQ:] \emph{Which are the mechanisms for avoiding and minimizing
    cybersecurity and misconfiguration issues in a CI container-based
    deployment system?}
\end{description}

Previous sections mentioned some of the common strategies used to
solve CI issues associated to this RQ;
mainly focused on the Image Generator step.
Hence,
it is possible to present a subset of scenarios for identifying \SD
validity.
Some of the answers extracted from this work are:

\begin{itemize}

    \item Even though is commonly accepted that the CI workflow relies
    on DevOps engineers experience,
    it is necessary to avoid unaware behaviors using a transparent and
    automated mechanism such as \SD.
    
    \item \SD,
    which works as an application firewall,
    has no impact compared with regular Docker use.
    
    \item Using a deployment engine based on YAML configuration files
    minimizes unaware deployments,
    simplifies repetitively tasks and makes more comprehensible
    automated monitoring process.  
    
    \item \SD allows to track and audit all commands sent to Docker.
    Its logging capabilities could be used as a tracking system having
    in mind the timestamp.
    
\end{itemize}

These points summarize the goal of the work presented and,
at the same time,
provide a concise and clear way to answer the RQ posed.

\subsection{Software Perspective}
\label{subsec:discussion:sw_perspective}

\SD offers many potential benefits regarding the CI process.
Some of these are:

\begin{itemize}

    %   Publicly available
    %\item SecDocker is released as an open source tool under MIT license. It is presented in a way that facilitates it deployment on the DevOps community. Language and libraries used for its use are based in of-the-shelf and open source software products. It is based on extended language Go and provides the interface to mainstream containerization solution: Docker.
    \item Publicly available.
    It is an open source tool released under the MIT license.
    The tool is presented in a way that makes deployment easier for
    the DevOps community.
    It is written in Go,
    a popular programming language,
    and offers a middleware solution for Docker,
    a mainstream containerization solution.
    
    %   (Configuration vs Simplicity)
    %\item 2. Flexibility and scalability:  DevOps and CI engineers should recognize that current release of SecDocker brings simplicity to the deployer process in Continuous Integration Process. Supported on a single configuration file, it is easy to define the farm of machines required for specific purposes fixing last minute/runtime missconfiguration issues.
    \item Flexibility, scalability and security.
    It should be noticeable from DevOps and CI engineers that the
    current release of \SD brings simplicity to CI and CD processes.
    Likewise,
    relying on different configuration files,
    makes easier to define all the requirements needed for an
    infrastructure and thus prevent misconfiguration issues
    related to last minute fixes,  to reduce performance issues associated with 
    lack of hardware resources or even software incompatibilities between versions.

    % Installation costs 
    %\item  Installation costs: The process of download, compiling and deployment is performed with three commands. It is documented in the Readme.md of the repository. 
    \item Installation costs.
    The process of downloading,
    compiling and deploying is performed with exactly three commands;
    as indicated in the documentation available in the GitHub repository.

    % - 3. Security () Developers
    % - SecDocker debería minimizar los efectos de malas configuraciones en los     contenedores utilizados.
    % - Malas configuraciones generan issues desde dos perspectivas, internas, para el equipo de devops y externas, en posibles entornos de despligue continuo.
    %\item Security \SD minimizes the risk of those accidental and frequent failures associated to misconfigurations when using containers  in CI and the number increases. The characteristics of \SD allow to minimize  performance issues, lack of hardware resources, or software incompatibility between versions. 
   % \item Security for DevOps Engineers.
    %\SD minimizes the risks of those accidental and frequent failures     associated to misconfigurations when using containers in CI.
%    Its features also allow to reduce performance issues,
 %   lack of hardware resources or even software incompatibilities between versions.

    % - Assurance(Speed vs Certainty):
    %\item  \SD users do not need to consider the trade-off  between speed and certainty.  Results presented in section\ref{subsec:experiment:performance} show similar performance and negligible differences when using Docker with or without \SD. 
    \item Assurance.
    \SD users do not need to consider the trade-off between speed and certainty.
    Results presented in Section~\ref{subsec:experiment:performance} show similar
    performance and negligible differences when using Docker with or without
    \SD.

\end{itemize}

% \subsection{Limitations}

However,
\SD also has certain shortcomings,
including:

\begin{itemize}

    %\item The solution do apply in the deployment phase, it does not cover previous steps. However, \SD architecture favors the use of plugins (Anchore, Notary) in order to offer this features. 
    \item The solution is only applicable to the deployment part of a
    CI/CD workflow;
    it does not cover previous steps.
    However,
    \SD architecture favors the use of plugins
    (like Anchore and Notary)
    in order to support such features.
    
    %\item \SD works as an application proxy, every time a client makes a Docker request, \SD intercepts the request and checks the IP and port specified, that needs to be the one associated to Docker. Currently, \SD does not enroute to other place that would add a new level of security allowing to hide connection elements to the user.
    \item \SD works as an application proxy.
    Each time a client makes a Docker request,
    \SD only intercepts it and checks its IP and port
    (which need to be the ones associated to Docker).
    Currently,
    \SD does not route these packages;
    which,
    on the other hand,
    would add a new level of security allowing to hide connection
    elements to the user.
    
    %\item The image provided in \SD configuration file is not validated. \SD does not check if the image provided by Docker Server is legit.
    \item The image provided in \SD's configuration file is not
    validated.
    More precisely,
    \SD does not check if the image provided by the Docker server
    is legit.
    
    %\item There are some unusual launch parameters that are not checked, such as those related to DNS or Input/Output. 
    \item Unusual launch parameters
    (like those related to DNS or Input/Output)
    are also not checked by \SD.
    
    %\item Once the container is running, \SD does not perform examination and provision of objective evidence that the container is running under the defined specifications or it is being used for intended purpose. 
    \item Once a container is running,
    \SD does not perform additional actions to test whether such
    container is executing under the defined specifications or
    is being used for the intended purpose.
    
\end{itemize}

%\subsection{Software metrics}%
%\label{subsec:experiment:metrics}

%Besides it is necessary to present some software metrics for offering  an assessment of the software developed in this study. %%   Short description of the experiment
%% This section explores SecDocker software metrics and its performance.
%On the one hand,
%metrics were measured against version XXX of the application and using
%SonarQube\footnote{https://www.sonarqube.org} and
%Golint\footnote{https://github.com/golang/lint} as code quality tools.

%%Particularly, it is a quantitative measurement study of SecDocker that will help top understand performance and efficiency of the application.

%%   Metrics presentation
%Formal code metrics for defining Maintainability and code quality of current version of our solution give details about how easy is to debug, maintain, and integrate new functionalities to \SD. 

%\SD has 1834 lines of code (LOC) and it has ana accummulative cyclomatic complexity of 206, mainly bounded in differennt functions of four files  tcpintercept (tcpproxy.go),  docker (security.go), commandline (command.go) and httpserver\_test (server\_test.go).

%The test metrics present a total number of 35 test cases, aggregated in 13 test functions (grouped by Table Driven Test)  and 87\% of test coverage.
 
%Finally, regarding its code quality,
%Golint detects 31 issues
%(28 related to naming and comments and 3 to coding structures)
%while SonarQube detects only 12 code smells and no bugs,
%vulnerabilities nor security hotspots.

%%%%%%%%%%%%%%%%%%%%%%%%%%%%%%%%%%%%%%%%%%%%%%%%%%%%%%%%%%%%%%%%%%%%%
Finally,
software metrics are presented in order to provide some sort of
assessment to the tool described in this paper.
These metrics can be used to define its maintainability and code quality
but also can give details about how easy is to debug,
maintain or integrate new functionalities to it.
Moreover,
they were measured against version v0.1-beta of the application and
using SonarQube\footnote{https://www.sonarqube.org} and
Golint\footnote{https://github.com/golang/lint} as code quality tools.

Considering the above,
\SD has 1834 lines of code
(LOC)
and an accumulative cyclomatic complexity of 206,
distributed among the different functions of four files:
\texttt{tcpintercept} (\emph{tcpproxy.go}), 
\texttt{docker} (\emph{security.go}),
\texttt{commandline} (\emph{command.go})
and \texttt{httpserver\_{}test} (\emph{server\_{}test.go}).
In addition,
it has a total number of 35 test cases---aggregated in 13 test functions
that are grouped by table-driven tests---and a 87\% of test coverage.
Lastly,
and regarding code quality,
Golint detects 31 issues
(28 related to naming and comments and 3 to coding structures)
while SonarQube detects only 12 code smells and no bugs,
vulnerabilities nor security hotspots.

% -------------------------------------------------------------------

\section{Conclusions}%
\label{sec:conclusions}

In conclusion,
it is important to harden CI workflow.
We knew from previous experiences that corporations refuse to deploy
new tools given the cost associated
(training, deployment, etc).
Thus,
the idea of providing a firewall app that allows maintaining the current
workflow was a key for designing \SD.

% By developing this application we have learned the possible threats a system with Docker can lead to. 
It is critical for every DevOps engineer to secure as much as possible their containers platforms. By developing \SD we have learned the possible threats of a CI system running containers, in particular the mainstream tool Docker. Performing a close analysis of the user input and harden the systems to minimize the possible attack surface and the capabilities the users can access to.

% -------------------------------------------------------------------

\begin{acknowledgements}
This work has been partially funded by the
``Universidad de León-Instituto Nacional de Ciberseguridad
(INCIBE)
Convention Framework about \emph{Detection of new threats and unknown
patterns}''
(Spain).
\end{acknowledgements}

% -------------------------------------------------------------------

\section*{Conflict of interest}

The authors declare that they have no conflict of interest.

% -------------------------------------------------------------------

\bibliographystyle{spmpsci}
\bibliography{references}

\end{document}